# Capillary pressure of van der Waals liquid nanodrops

Roumen Tsekov and Borislav V. Toshev
Department of Physical Chemistry, University of Sofia, 1164 Sofia, Bulgaria

The dependence of surface tension on a nanodrop radius is important for the new-phase formation process. It is demonstrated that the famous Tolman formula is not unique and the size-dependence of the surface tension can distinct for different systems. The analysis is based on a relationship between the surface tension and disjoining pressure in nanodrops. It is shown that the van der Waals interactions do not affect the new-phase formation thermodynamics since the effects of the disjoining pressure and size-dependent component of the surface tension cancel each other.

In small systems all thermodynamic parameters depend on the system size. Their values differ from those in large systems, which are usually given in thermodynamic reference books. For instance, if an interface approaches another one the transition phase regions overlap. As a result, both the surface tension and pressure in the gap dividing the two surfaces change, due to onset of disjoining pressure. The present paper focuses on small droplets, where the surface tension depends on the drop radius. The description of such a system is important for the theory of nucleation [1-3], smog, fog and nano-emulsion stability, etc.

Let us consider a globular mass surrounded by a bulk gas phase, which is sufficiently small that no region of the mass can be regarded as homogeneous phase owning bulk properties. The well-known Tolman formula [4] for the effect of drop size on the surface tension [3]

$$\sigma = \sigma_\infty (1 - 2\delta_\infty / R) \qquad (1)$$

is derived by careful analysis of the Gibbs remarks on the problem [5]. Here $\sigma$ and $\sigma_\infty$ are the surface tension of a drop and a planar liquid-vapor interface, respectively, $\delta_\infty$ is the so-called Tolman length, and $R$ is the drop radius. In the present paper the latter is considered to mark the place of the surface of tensions. However, some assumptions in the derivation of Eq. (1) are considered to be questionable [6] and this is corroborated by the fact that the linear dependence of $\sigma$ vs. $1/R$ has not been experimentally verified. Moreover, neither the magnitude nor even of the sign of the Tolman length is obvious. Thus, the size-dependence of the surface tension is still an open area of research and several new papers in this field have been published [7-14], amongst them articles with computer simulations.

The problem of the size-dependent surface tension can be considered from an alternative perspective. The thin liquid film is another small system, where the two surfaces of discontinuity overlap and no bulk liquid core within the film exists. This fact is quantitatively manifested via the film disjoining pressure $\Pi$ introduced by Derjaguin [15]. The present study aims to calculate the disjoining pressure for a drop and then, using the relation between $\Pi$ and $\sigma$, to obtain the size-dependence of the nanodrop surface tension. In what follows we consider the thermodynamic description of a small drop and a definition of the disjoining pressure of the drop is derived. As an example, the drop van der Waals disjoining pressure is calculated via the method of Hamaker [16] with the equation of London [17].

The mechanical behavior of a drop is determined by its pressure tensor. The normal component $P_N$ of the pressure tensor in the drop center differs from the pressure $P_G$ in the homogenous gas phase and their difference is given by the hydrostatic Laplace law

$$P_N - P_G = 2\sigma / R \tag{2}$$

which holds in any case, no matter if there is a bulk liquid core within the drop or the drop is so small that no part is homogeneous. For the latter case, according to the Gibbs consideration [5], a hypothetical liquid phase, having the same temperature and chemical potential as the gas phase, is attributed to the mass, which is conceived as existing within the dividing surface. The pressure $P_L$ in this liquid phase differs from $P_N$ and their difference is known as disjoining pressure [14, 2]

$$\Pi \equiv P_N - P_L \tag{3}$$

According to thermodynamics, the characteristic function of a system at constant temperature and chemical potential is the omega potential. The $\Omega$ potential of a liquid drop is given by the expression

$$\Omega = -P_L V + \sigma O \tag{4}$$

where $V = 4\pi R^3 / 3$ and $O = 4\pi R^2$ are the drop volume and surface area, respectively. Note that according to the Gibbs-Duhem relation for the liquid phase the pressure $P_L$ is constant at constant temperature and chemical potential. On the other hand the work necessary to change the drop size can be calculated via the relation

$$d\Omega = -P_N dV + \sigma dO \tag{5}$$

Substituting $\Omega$ from Eq. (4) and employing definition (3) one yields the Gibbs-Duhem relation for a drop at constant temperature and chemical potential

$$\Pi dV = -Od\sigma \tag{6}$$

For the case of a spherical drop Eq. (6) can be easily transformed to the following expression

$$\Pi = -\partial\sigma / \partial R \tag{7}$$

which relates the disjoining pressure and the first derivative of the surface tension with respect to the drop radius. Equation (7) states that the dependence of the surface tension on the drop size indicates excess energy in the small drop compared to the bulk phase. This expression is very similar to $\Pi = -2\partial\sigma/\partial h$ for foam films [18, 2], where $h$ is the film thickness. Combining now Eqs. (2) and (3) one can calculate the capillary pressure of the drop

$$P_L - P_G = 2\sigma/R - \Pi = 2\sigma/R + \partial\sigma/\partial R \tag{8}$$

The last expression is obtained by employing Eq. (7) and represents a known result from the literature.

Finally, by direct integration of Eq. (7) a useful expression for calculating the surface tension of a drop is derived

$$\sigma = \sigma_\infty + \int_R^\infty \Pi dR \tag{9}$$

Equation (9) shows that the size-dependence of the surface tension is not universal in contrast to the Tolman formula. The surface tension is determined by the specific interactions within the drop, which reflect various components of the disjoining pressure. In the following part, the contribution of the most widespread force, i.e. the van der Waals attraction, is considered. The method of Hamaker [16] is used to calculate the disjoining pressure of the drop. Thus, the van der Waals disjoining pressure is equal to the energy per unit volume of the interaction of a molecule, placed in the center of a bubble with radius $R$, with the surrounding liquid. Using the London equation [17] for the energy of attraction between two molecules, the following expression is obtained

$$\Pi_{VW} = -\frac{1}{v_L^2}\int_R^\infty \frac{\lambda}{r^6} 4\pi r^2 dr = -\frac{4A}{3\pi R^3} \tag{10}$$

where $\lambda$ is the London constant, $v_L$ is volume per molecule of liquid, and $A = \pi^2 \lambda / v_L^2$ is the Hamaker constant. By introducing Eq. (10) into Eq. (9) and integrating the result one yields

$$\sigma = \sigma_\infty (1 - a^2 / R^2) \tag{11}$$

where $a^2 \equiv 2A / 3\pi\sigma_\infty$. For the typical values of the Hamaker constant and surface tension the length $a$ is of the order of nanometers and, hence, deviations of $\sigma$ from $\sigma_\infty$ are essential in nanodrops only. It is evident that the size-dependent surface tension from Eq. (11) is inversely proportional to square of the drop radius, and differs from the Tolman formula (1). Also, according to Eq. (11) the surface tension of a nanodrop is always lower than $\sigma_\infty$ on a flat surface.

By introducing the surface tension from Eq. (11) in Eq. (8), one obtains the following expression for the capillary pressure in a van der Waals liquid drop

$$P_L - P_G = 2\sigma_\infty / R \tag{12}$$

This expression holds either for relatively large drops, where $P_L = P_N$, or for very small drops, where $P_L \neq P_N$. In both cases the thermodynamic Laplace law (12) involves the surface tension $\sigma_\infty$ of a plane interface. The authors are of the opinion that Eq. (12) is in fact expressed by the Gibbs statement: "With this understanding with regard to the phase of the fictitious interior mass, there will be no ambiguity in the meaning of any of the symbols which we have employed, when applied to cases in which the surface of discontinuity is spherical, however small the radius may be" [5]. Combining Eqs. (8) and (12) the following equation can be obtained

$$2\sigma_\infty / R + \Pi_{VW} = 2\sigma / R \tag{13}$$

This equation can also be derived by the force balance method, in which a spherical drop with radius $R$ is cut in two equal parts and one of these parts is "solidified" (the Steven method). The liquid part of the drop is characterized by $\sigma$ and $P_N$, while the "solidified" part of the drop is represented by $\sigma_\infty$ and $P_L$. Hence, at mechanical equilibrium the force balance

$$\pi R^2 P_N - 2\pi R \sigma = \pi R^2 P_L - 2\pi R \sigma_\infty \tag{14}$$

is applicable, which also results in Eq. (13). A similar situation exists in the mechanical description of the transition zone between a plane-parallel liquid film and a bulk liquid meniscus. In this case, the Derjaguin approach [19] with constant surface tension and disjoining pressure and the de Feijter approach [20] with size-dependent surface tension without disjoining pressure are found to be equivalent [21]. However, Eq. (12) holds only for van der Waals liquids. It is shown that both these approaches are inappropriate for drops with electrostatic forces [22]. The correct solution requires either size-dependent surface tension or disjoining pressure.

The research presented in this paper shows that there is a general relationship between the surface tension of a drop and the disjoining pressure. Since the latter reflects many different types of interactions, e.g. van der Waals, electrostatic, hydrophobic, structural, etc., the dependence of the surface tension on the drop radius is not universal. It is an integral effect of all interactions in the drop and, consequently, is system specific. For instance, according to Eq. (11) the van der Waals effect is inversely proportional to the square of the drop radius [23]. It has been demonstrated in this paper that the van der Waals interactions do not affect the equations describing the thermodynamics of new-phase formation.